# All-optical switching in dye-doped DNA nanofibers

Adam Szukalski,[a,b] Maria Moffa,[b] Andrea Camposeo,[b] Dario Pisignano,[b,c] and Jaroslaw Mysliwiec [a]

[a.] *Faculty of Chemistry, Wroclaw University of Science and Technology, Wybrzeze Wyspianskiego 27, 50-370 Wroclaw, Poland*
[b.] *NEST, Istituto Nanoscienze-CNR, Piazza S. Silvestro 12, I-56127 Pisa, Italy*
[c.] *Dipartimento di Fisica, Università di Pisa, Largo B. Pontecorvo 3, I-56127 Pisa, Italy.  E-mail: dario.pisignano@unipi.it*

All-optical switches are introduced which are based on deoxyribonucleic acid (DNA) in the form of electrospun fibers, where DNA is semi-intercalated with a *push-pull*, luminescent nonlinear pyrazoline derivative. Optical birefringence is found in the organic nanofibers, with fully reversible switching controlled through continuous-wave laser irradiation. The photoinduced signal is remarkably large, with birefringence highlighted by optically-driven refractive index anisotropy approaching 0.001. Sub-millisecond characteristic switching times are found. Integrating dye-intercalated DNA complex systems in organic nanofibers, as convenient and efficient approach to template molecular organization and controlling it by external stimuli, might open new routes for realizing optical logic gates, reconfigurable photonic networks and sensors through physically-transient biopolymer components.

1. **Introduction**

The nonlinear optical properties of bulk organic materials and micro-architectures based on them are generally associated to their response far from resonance, including refractive index changes, which originates mostly from the electronic cloud deformation occurring at molecular scale. Indeed, the potentially large, fast nonlinear behaviour of various organic systems is related with π-electron system delocalization[1] and is non-resonant in its nature. However, a large enhancement of the nonlinear behaviour might occur when excitation light is close to resonance, with response timescales up to $10^{-2}$ s. Importantly, additional control of the nonlinear optical properties of organic systems is provided by the materials embedding the photoactive molecules. Examples of matrix compounds which can host organic molecules relevant in this framework include biopolymers like deoxyribonucleic acid (DNA),[2-4] possibly functionalized by different types of surfactants,[5,6] starch,[7] or collagen.[8]

In particular, DNA which has been studied almost exclusively by biologists over the past half-century,[9] is now a well-established polymer for developing a variety of processes and devices in photonics and nanotechnology.[9,10] It was shown that DNA modified with surfactant molecules, which is insoluble in water but easily processable in organic solvents, leads to optical materials with significant transparency in the visible range.[11,12] Also, doping such a matrix with light-emitting molecules changes its properties and might allow blends to be obtained which show amplified spontaneous emission,[13] lasing[14] or random lasing.[15] Moreover, doping with photochromic molecules makes the resulting composite suitable for fast holographic recording due to the dye semi-intercalation.[12] Modified DNA complexes are being also used in organic electronics to realize transistors, or light-emitting diodes.[16-25] The by far largest amount of works applying these systems in photonics and electronics are based on film structures. However, the random and variable distribution of dye molecules, which is generally achieved in films, might prevent the effective and coherent addressing of individual molecules or ensembles made of them, which is highly important in order to exploit molecular functions in practically usable devices. In this respect, integrating dye-intercalated DNA complex systems in micro- and nanostructures could open new possibilities for studying their photo-physical properties, for managing them in a coherent way, and for templating their organization making the modification of molecular status by external stimuli much more convenient and efficient, also due to the greatly enhanced surface-to-volume ratio. Other important aspects are related to the low-cost and potential easy processability of DNA-based nanostructures, especially nanofibers. In these systems, thanks to the unique double helix structure and interaction between each host DNA chain and low-molar mass dyes (guests), fully reversible and stable all-optical bio-switches might be feasible. So far, various nonlinear mechanisms have been introduced for all-optical switching of light beams, such as saturable absorption and optical limiting,[26-30] nonlinear refractive index and photoinduced birefringence,[6,31,32] thermo-optic effects,[33] using organics or 2D nanomaterials either in liquids or in films and nanosheets. Nevertheless, switches based on organic nanofibers are still poorly studied, and optical properties, such as refractive index, have never been controlled by external light, which is critically important for all-optical interconnects.

Here, we demonstrate all-optical switching in organic fibers consisting of DNA functionalized with cetyltrimethylammonium chloride surfactant (DNA-CTMA) and doped with a nonlinear pyrazoline derivative chromophore (PY-$p$CN) as typical *push-pull* molecular system (Fig. 1a). Fibers and excitation laser light serve as templates for molecular ordering leading to optical anisotropy and driving the system into two well-defined states used for the switch construction. Based on the achieved light-controlled birefringence, a great potential is suggested for these materials for fast all-optical switching applications. These might include remotely-reconfigurable photonic networks and sensors,[34] opto-logic gates and multiplexers,[35,36] and physically-transient optoelectronic components[2] based on natural biopolymers.

2. **Results and discussion**

Previous works found that the specific interaction between the DNA and photochromic azo-molecules (semi-intercalation) leads to a fully-reversible photoisomerization which is much faster than that of typical host-guest systems.[6,37-42] Thanks to this unique structure promoted by the bio-organic matrix (Fig. 1b), it is possible to obtain semi-hooking of the dye molecules and to straightforwardly control their spatial arrangement by polarized light irradiation, thus causing dynamic and fully-reversible modulation of refractive index. The so-achieved, photo-controlled system features two well-defined and reversible states. PY-*p*CN, which is a *push-pull*, π-conjugated compound featuring intramolecular charge transfer, was achieved by the synthetic route shown in Fig. 1c and detailed in the Supplementary Information file (molecular characterization are reported in Figure S1 and S2).[42] Recently, pyrazoline derivatives were utilized by us for light amplification (in laser action or random lasing),[3,43-45] multiphoton absorption,[42,46] generation of light at higher harmonics,[47,48] and their structure was found to be modified by light in a reversible way through *trans* or *cis* states.[49,50]

Electrospun PY-*p*CN/DNA-CTMA mats consist of mainly ribbon-shaped microfibers (Fig. 2a) with roughly bi-modal distributions in terms of transversal size peaked at about 250 nm and at about 1.2 μm, respectively (Fig. S3). The absorption spectrum of the active material, with the laser lines used for pump and probe experiments is presented in Fig. S4 in the Supplementary Information. In our fibers, the overall degree of dopant alignment can be easily controlled by linearly polarized laser light, thus leading to refractive index anisotropy.

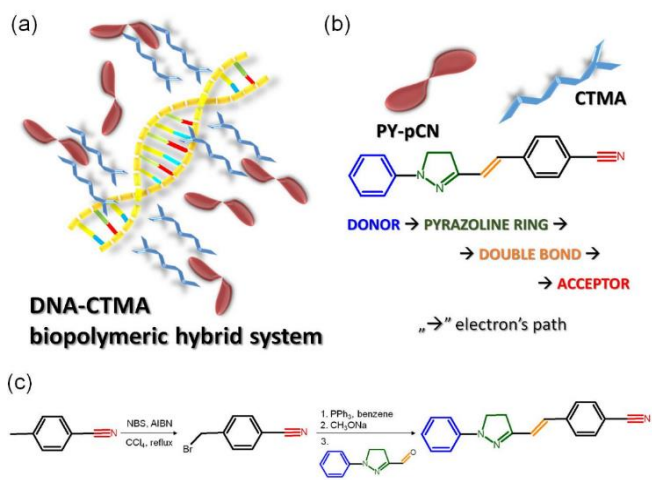

**Figure 1.** (a) Scheme of the photoisomerizable dye semi-intercalation in the DNA-CTMA matrix. (b) Chemical structure of photoactive medium (PY-*p*CN). Arrows in the schematics indicate intramolecular charge transfer paths for electrons. (c) Steps of PY-*p*CN synthesis. NBS: N-Bromosuccinimide; AIBN: Azobisisobutyronitrile; PPh$_3$: Triphenylphosphane.

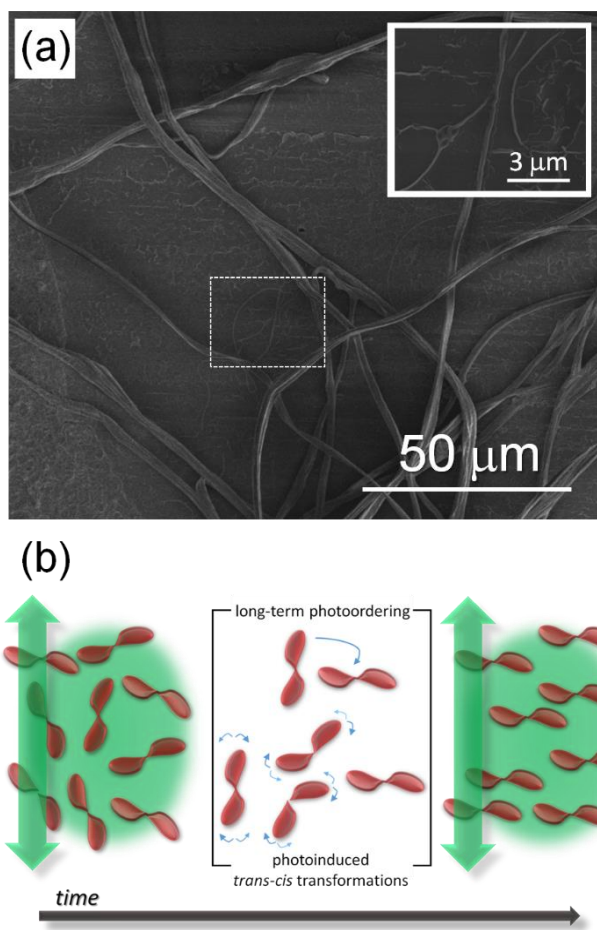

**Figure 2.** (a) SEM micrograph of PY-*p*CN-doped DNA-CTMA fibers. Inset: higher magnification view of the region highlighted by the dashed rectangle, showing fibers with diameters down to >100 nm. (b) Scheme of the photoinduced optical birefringence generated through molecular alignment in electrospun nanofibers, induced by linearly polarized laser light. Vertical green arrows: direction of light polarization; single blue arrow: long-term photo-ordering; doubled blue arrows: photo-induced *trans-cis-trans* transitions.

Indeed, the fiber-embedded photoswitchable molecules can be oriented independently on the filament direction, the transition from the initially-isotropic system to an anisotropic one being promoted by laser excitation. Two regimes can be distinguished in this respect, namely short- and long-term photo-ordering, as schematized in Fig 2b.

In the short-term ($10^{-9}$ to $10^{-3}$ s),[51] proper photoinduced molecular transitions like *trans-cis-trans* photoisomerizations should be considered. For this process to take place, the excitation wavelength should be within the absorption range of the active medium, and there should be enough free volume in the matrix to enable the involved transition. In parallel, thermal relaxation occurs converting the metastable *cis*-form to the stable *trans*-isomer, and continuously leading the system towards a more isotropic arrangement (i.e. to a lowest-energy thermodynamic state) over timescales typically ranging from $10^{-3}$ s to a few minutes.[51-53] The onset of significant photo-alignment at this timescales can be highlighted modulating the pump beam at frequencies above a few hundreds of Hz, and measuring the refractive index anisotropy by using an oscilloscope in AC mode in pump-probe experiments (Fig. S5). Following a very large-number of so-defined photoinduced *trans-cis-trans* transitions, a photostationary state is achieved and long-term photo-ordering is established (highlighted by DC mode in the detecting oscilloscope), with in principle all the dopants oriented perpendicularly with respect to the laser light polarization direction and timescales from seconds to days. This is promoted by rotational diffusion[54] of the nonlinear chromophores as a consequence of the coupling with their local microenvironment (generally constituted by the solvent for solutions, and by the polymer matrix for solid-state samples). The photoinduced birefringence measured in PY-*p*CN/DNA-CTMA fibers and its dependence on the excitation intensity ($I$) are shown in Fig. 3 for long-term photo-ordering. A clear linear behaviour is found for the photoinduced refractive index anisotropy ($\Delta n$), i.e. $\Delta n(I) = n_2 I$ with $n_2 = 2.0 \times 10^{-10}$ m$^2$/W, without evidence of signal saturation in the investigated range of pump intensities. This indicates remarkable photostability of the realized optical material. By increasing the intensity of the pumping beam up to about 4 W/cm$^2$, the refractive index anisotropy, $\Delta n$, approaches values of the order of 0.001 (Fig. 3b).

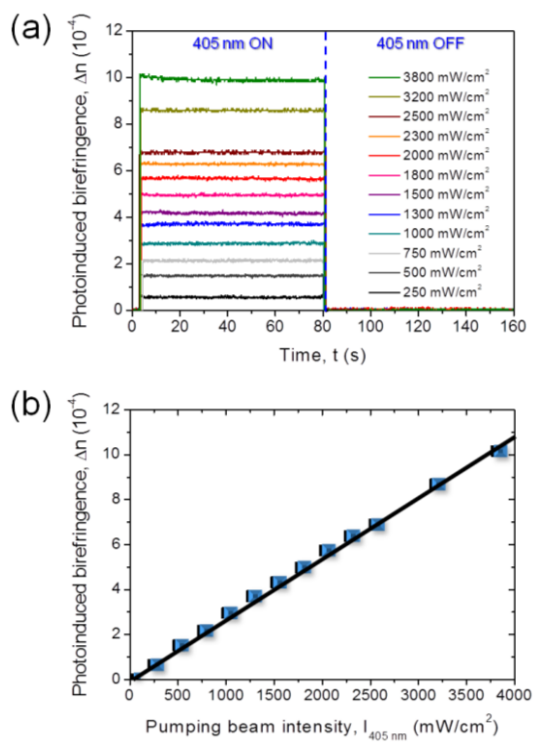

**Figure 3.** Dependence of measured photoinduced birefringence on power excitation intensity from the pump beam (long-term photo-ordering). (a) Switching behaviour. (b) Dependence of the photoinduced refractive index anisotropy (birefringence), $\Delta n$, on the pumping beam intensity.

Similarly to optical Kerr effect, the intensity-induced linear birefringence can be described by an effective third-order nonlinear optical susceptibility,[55-57] $\chi_{eff}^{(3)} \cong n_2 n_0^2 \varepsilon_0 c$, where $n_0$ denotes the refractive index of the system without excitation from the pump beam ($n_0$=1.5),[58] $\varepsilon_0$ is the dielectric constant and $c$ the light velocity, both in vacuum conditions. In electrospun PY-*p*CN/DNA-CTMA fibers, the third-order nonlinear optical susceptibility was estimated to be of the order of $1\times10^{-12}$ m$^2$/V$^2$, in line with those found in well-established organic systems embedding nonlinear chromophores.[49,50]

As unprecedented evidence of such all-optical switching effect in electrospun organic fibers, these findings make such systems a promising alternative against spiropyran/merocyanine,[59] film-embedded azobenzene chromophores,[52,60] polymethine compounds,[61,62] and conjugated polycyanine.[63] Exemplary, monoexponential increase and decay behaviours for the long-term photo-ordering in the fibers are shown in Fig. S6, indicating symmetric response during switching 'on' and 'off' the system, with characteristic time for achieving the stationary state of photo-ordering of about 0.4 ms. In Fig. 4a we present the photoinduced birefringence acquired with the oscilloscope in DC mode and a pump beam modulator. The remarkable signal stability at long times is highlighted in Fig. 4b. Even upon multiple photoinduced birefringence processes in the PY-*p*CN/DNA-CTMA fibers, Δ*n* can be re-build quickly and, what is crucial for realizing optical switching devices, without appreciable efficiency losses.

To rationalize the molecular switching process more in depth and investigate the range of switching frequency accessible through electrospun PY-*p*CN/DNA-CTMA nanofibers, we investigated the correlation between photoinduced birefringence and laser light modulation frequency (Fig. 5a). When the modulation frequency is roughly 300 Hz or lower, laser irradiation intervals are long enough to approach the photostationary state and a long-term optical anisotropy generation. This is evidenced by the stable values of the photoinduced birefringence measured for pump modulations in the range of frequency 50-300 Hz (Fig. 5a). For faster modulations, the maximum photoinduced birefringence decreases, a result suggesting a predominant role of *trans-cis-trans* photoisomerization and thermal relaxation against long-term photo-ordering. For modulation at 800 Hz, a Δ*n* of 0.00017 is found, highlighting a switching behaviour still significant upon approaching kHz frequencies (Fig. 5b). The fast-responding system shows increase and decay times $\tau_{AC}$(inc) = $\tau_{AC}$(dec) = 0.35 ms at a modulation frequency of 200 Hz (Fig.s 5c and 5d). Such characteristic switching times are an order of magnitude lower than the ones observed for the PY-*p*NO$_2$ embedded in PMMA.[49] Hence, the nanostructured biopolymer matrix here developed allows sub-millisecond all-optical switching to be achieved. This is relevant for various spectroscopic, imaging and manufacturing applications, where either intensity or phase modulation of light beams is required.[60,64]

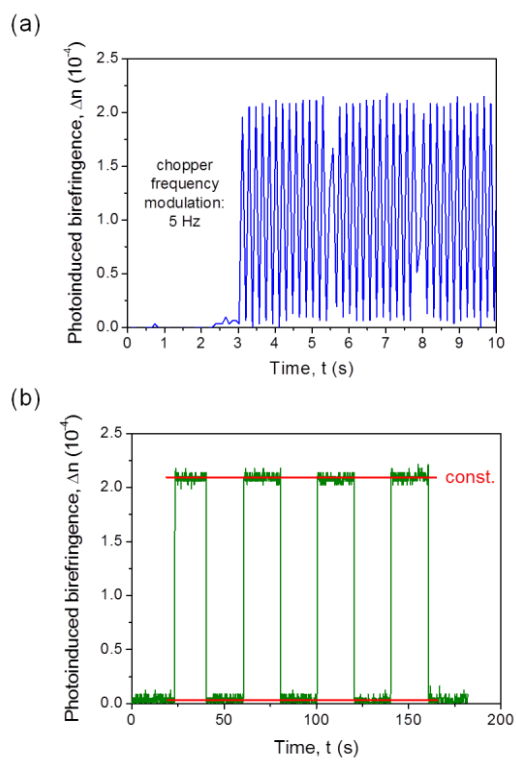

**Figure 4.** (a) $\Delta n$ values measured upon applied light modulation and (b) multiple photoinduced birefringence process in PY-*p*CN/DNA-CTMA fibers ($I_{pump}$ = 750 mW/cm$^2$).

The realization of DNA fibers by electrified jets,[65] promoting orientation of macromolecules within individual bio-polymer filaments, might be especially relevant in stabilizing photo-induced optical anisotropy. Indeed, the higher molecular order induced by jet elongation during electrospinning, leading to fibers with polymer chains prevalently oriented along the longitudinal axis of the filaments, directly favour the rise of optical anisotropies,[66-68] a property typically enhanced in thinner electrospun fibers.[68] The complex interplay of these mechanisms with the photo-alignment of the initially-isotropic embedded chromophores might make dye-doped DNA fibers efficient, repeatable and fast-responsive optical switches, with remarkable non-linear optical signals and stability. Moreover, the possibility of controlling the molecular order in electrospun nanofibers by processing parameters and ultimately through the fiber transversal size[68] could constitute an additional variable for tailoring the nonlinear optical properties of embedded compounds. This approach can be pretty useful for photonic computing architectures, where logic gates are required to control data transfer. For these applications, functionalized DNA in anisotropic template nanostructures appears unique in its association with semi-intercalation of nonlinear chromophores, and highly effective for supporting molecular photoalignment.

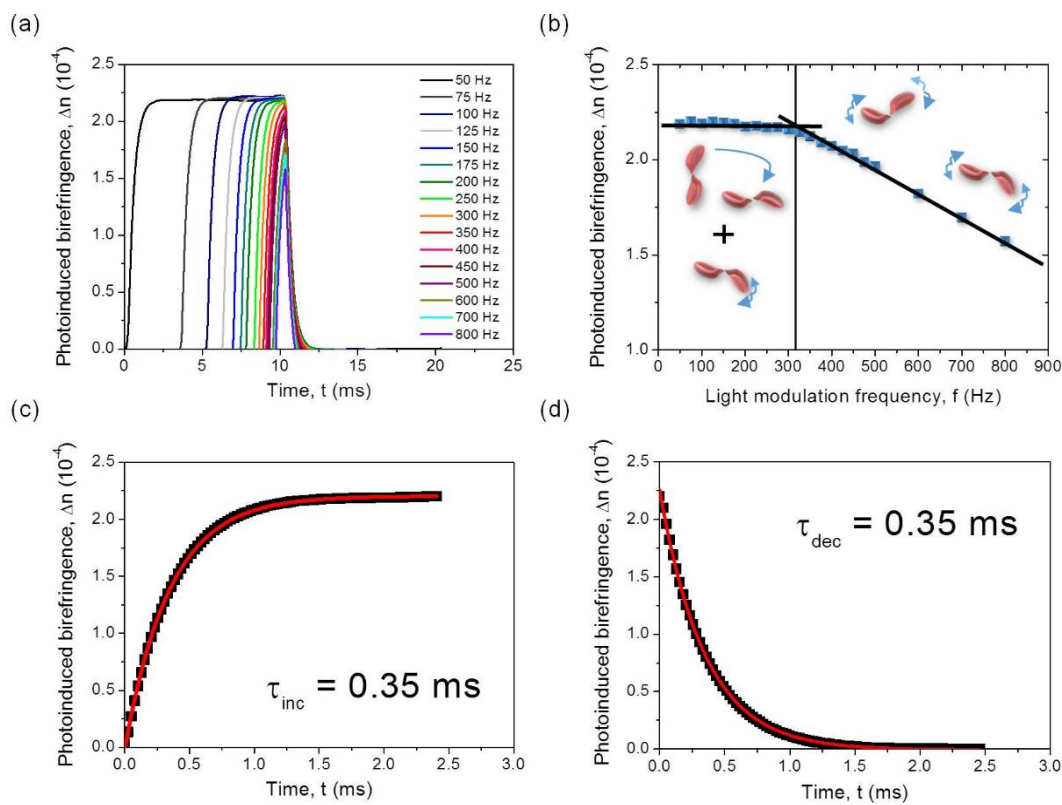

**Figure 5**. Photodynamic changes of the induced birefringence in PY-*p*CN/DNA-CTMA fibers. (a) AC component of the photoinduced birefringence signal for various modulation frequency of the pump light. (b) Δ*n* vs. modulation frequency. The vertical continuous line highlights the transition from long-term (modulation frequency < 300 Hz) to short-term photo-ordering (modulation frequency >300 Hz)- Lines superimposed to experimental data are guides for the eye. Corresponding birefringence increase (c) and decrease (d), with exponential fits of experimental results. Applied modulation frequency: 200 Hz. $I_{pump}$ = 750 mW/cm$^2$.

3. **Materials and methods**

**Materials.** We used DNA functionalized with a CTMA surfactant (Sigma-Aldrich) and doped with the luminescent and photoisomerizable chromophore (E)-2-(2-(1-phenyl-4,5-dihydro-1H-pyrazol-3-yl)vinyl)benzonitrile (PY-*p*CN) (Fig. S2). For realizing PY-*p*CN-doped DNA-CTMA fibers, the DNA-CTMA complex was dissolved in butanol with 4% dry mass weight ratio (w/w) and stirred for 24 h. A second solution was obtained using the same solvent and PY-*p*CN at 2% (also w/w) dye concentration. A mixture was then obtained by proper quantities of the two solutions to reach a 1% concentration of nonlinear chromophore in the dry mass of biopolymer, which was electrospun using a feeding rate of 0.5 mL/h, a distance between spinning tip and collector surface of 20 cm, and an applied voltage of 10 kV. The morphology of the doped fibers was investigated by scanning electron microscopy (SEM, Nova NanoSEM 450, FEI) with an accelerating voltage of 3 kV and an aperture size of 30 μm, following thermal deposition of 5 nm of Cr (PVD75, Kurt J. Lesker Co.). The average diameter of the fibers was calculated from the SEM micrographs using an imaging software.

**Spectroscopy.** All-optical switching processes were investigated by a pump-probe experimental set-up (Fig. S5 in the Supplementary Information file), described in detail elsewhere.[49,50] Briefly, we used two CW laser light sources with 405 nm and 633 nm wavelength to generate pump and probe beams, respectively. The two beams were converged at one point located on the sample. For characterizing chromophore photo-alignment at long time-scales (associated to the quasi-static component of the photoinduced refractive index anisotropy, $\Delta n$), the pump beam was switched "on" and "off" manually to promote optical anisotropy (pump laser ON) or let the system goes through thermal relaxation processes under darkness (pump laser OFF), respectively. A cross-polarized system was used, equipped with a sensitive photodiode combined with an oscilloscope in DC mode to acquire the above dynamics. For assessing the dynamic component of photoinduced birefringence, namely for characterization at shorter timescale, a mechanical chopper with controlled frequency modulation was embedded in the experimental set-up to modulate the pump intensity, and the analysing oscilloscope set in AC mode. The photoinduced refractive index anisotropy ($\Delta n$) is determined as a function of time ($t$) and excitation intensity ($I$) at the same time, and correlated with phase changes ($\Delta\varphi$, which is also function of $t$ and $I$), wavelength of the probe beam ($\lambda$) and sample thickness ($d$) as[49,50] $\Delta n(I,t) = \lambda \, \Delta\varphi(I,t)/2\pi d$. The correlation between the intensities of incident ($I_0$) and transmitted ($I_{trans}$) probe light involved with the phase change and resulting refractive index anisotropy is given by:[49,50] $I_{trans}(t) = I_0 \sin^2[\pi d \Delta n(I,t)/\lambda]$. This approach allows changes in the material at nanoscopic scales to be assessed by straightforward power measurement of the probe laser beam.

4. **Conclusions**

In summary, optical switches based on electrospun fibers made by properly-functionalized biological material have been proposed. DNA-CTMA doped with pyrazoline derivative features effective, stable, quick and fully-reversible modulations of the refractive index. Time constants associated with the involved dynamic changes (*trans-cis-trans* molecular transformations and long-term photoalignment) are $\tau_{AC}(inc) = \tau_{AC}(dec) = 0.35$ ms, respectively, suggesting these fibrous materials as valuable competitor of well-known azobenzene-based systems. The present optical switches based on a bio-matrix and pyrazoline derivative can be utilized for the design and development of efficient and fast sensing schemes, photonic logic operators, and organic multiplexers based on functional nanofibers.

## 5. Acknowledgements

A. S.: Supported by the Foundation for Polish Science (FNP). J.M.: This work was financially supported by The National Science Centre, Poland (2016/21/B/ST8/00468) and by statutory founds of the Wroclaw University of Science and Technology. The research leading to these results has also received funding from the European Research Council under the European Union's Seventh Framework Programme (FP/2007-2013)/ERC Grant Agreement n. 306357 ("NANO-JETS") (D.P.) and under the European Union's Horizon 2020 Research and Innovation Programme (Grant Agreement n. 682157, "xPRINT") (A.C.). D.P. also acknowledges the support from the project PRA_2018_34 ("ANISE") from the University of Pisa. Karolina Haupa is gratefully acknowledged for infrared spectroscopy on PY-$p$CN.

## 6. Notes and references

1   D. S. Chemla and J. Zyss, J. *Nonlinear Optical Properties of Organic Molecules and Crystals*, Academic Press, New York, 1987.

2   A. Camposeo, P. Del Carro, L. Persano, K. Cyprych, A. Szukalski, L. Sznitko, J. Mysliwiec and D. Pisignano, *ACS Nano*, 2014, **8**, 10893.

3   I. Rau, A. Szukalski, L. Sznitko, A. Miniewicz, S. Bartkiewicz, F. Kajzar, B. Sahraoui and J. Mysliwiec, *Appl. Phys. Lett.,* 2012, **101**, 171113.

4   A. J. Steckl, *Nat. Photonics,* 2007, **1**, 3.

5   E. M. Heckman, J. A. Hagen, P. P. Yaney, J. G. Grote and F. K. Hopkins, *Appl. Phys. Lett.,* 2005, **87**, 211115.

6   A. Miniewicz, A. Kochalska, J. Mysliwiec, A. Samoc, M. Samoc, and J. G. Grote, *Appl. Phys. Lett.,* 2007, **91**, 041118.

7   K. Cyprych, L. Sznitko and J. Mysliwiec, *Org. Electron.,* 2014, **15**, 2218.

8   K. Cyprych, M. Janeczko, I. Rau, F. Kajzar and J. Mysliwiec, *Org. Electron.,* 2016, **39**, 100.

9   R. A. Goodnow, *A Handbook for DNA-Encoded Chemistry: Theory and Applications for Exploring Chemical Space and Drug Discovery*, Wiley, New Jersey, 2014.

10  J. G. Grote, D. E. Diggs, R. L. Nelson, J. S. Zetts, F. K. Hopkins, F. K. Ogata, J. A. Hagen, E. Heckman, P. P. Yaney, M. O. Stone, and L. R. Dalton, *Mol. Cryst. Liq. Cryst.,* 2005, **426**, 3.

11  E. M. Heckman, J. A. Hagen, P. P. Yaney, J. G. Grote and F. K. Hopkins, *Appl. Phys. Lett.*, 2005, **87**, 211115.

12  C. A. Lazar, F. Kajzar, I. Rau, L. and A.-M. Manea, *Synth. Met.*, 2016, **221**, 120.

13  J. Mysliwiec, L. Sznitko, A. Miniewicz, F. Kajzar and B. Sahraoui, *J. Phys. D: Appl. Phys.*, 2009, **42**, 1.

14  Y. Kawabe and K.-I. Sakai, *Nonlinear Opt., Quantum Opt.*, 2011, **43**, 273.

15 L. Sznitko, A. Szukalski, K. Cyprych, P. Karpinski, A. Miniewicz and J. Mysliwiec, *Chem. Phys. Lett.,* 2013, **576**, 31.

16 Y.-W. Kwon, C. H. Lee, D.-H. Choi and J.-I. Jin, *J. Mater. Chem.,* 2009, **19**, 1353.

17 J. A. Hagen, W. Li, A. J. Steckl and J. G. Grote, *Appl. Phys. Lett.,* 2006, **88**, 171109.

18 D. Wanapun, V. J. Hall, N. J. Begue, J. G. Grote and G. J. Simpson, *ChemPhysChem,* 2009, **10**, 2674.

19 E. M. Heckman, J. G. Grote, F. K. Hopkins and P. P. Yaney, *Appl. Phys. Lett.,* 2006, **89**, 181116.

20 J. Zhou, Z. Y. Wang, X. Yang, C.-Y. Wong and E. Y. B. *Opt. Lett.,* 2010, **35**, 1512.

21 Z. Yu, W. Li, J. A. Hagen, Y. Zhou, D. Klotzkin, J. G. Grote and A. J. Steckl, *Appl. Opt.,* 2007, **46**, 1507.

22 M. Leonetti, R. Sapienza, M. Ibisate, C. Conti and C. López, *Opt. Lett.,* 2009, **34**, 3764.

23 T. Sasaki, H. Ono and N. Kawatsuki, *Jpn. J. Appl. Phys.,* 2007, **46**, 1579.

24 I. G. Marino, D. Bersani and P. P. Lottici, *Opt. Mater.,* 2001, **15**, 279.

25 Y. J. Wang and G. O. Carlisle, *J. Mater. Sci.: Mater. Electron.,* 2002, **13**, 173.

26 X. Zhang, S. Zhang, C. Chang, Y. Feng, Y. Li, N. Dong, K. Wang, L. Zhang, W. J. Blau, and J. Wang, *Nanoscale*, 2015, **7**, 2978.

27 J. Huang, N. Dong, S. Zhang, Z. Sun, W. Zhang, and J. Wang, *ACS Photonics*, 2017, **4**, 3063.

28 N. Dong, Y. Li, S. Zhang, X. Zhang, and J. Wang, *Adv. Optical Mater.*, 2017, **5**, 1700543.

29 Y. Chen, T. Bai, N. Dong, F. Fan, S. Zhang, X. Zhuang, J. Sun, X. Zhang, J. Wang, and W. J. Blau, *Prog. Mater. Sci.*, 2016, **84**, 118.

30 N. Dong, Y. Li, Y. Feng, S. Zhang, X. Zhang, C. Chang, J. Fan, L. Zhang, and J. Wang, *Sci. Rep.*, 2015, **5**, 14646.

31 N. Dong, Y. Li, S. Zhang, N. McEvoy, X. Zhang, Y. Cui, L. Zhang, G. S. Duesberg, and J. Wang, *Opt. Lett.*, 2016, **41**, 3936.

32 K. Wu, C. S. Guo, H. Wang, X. Y. Zhang, J. Wang, and J. P. Chen, *Opt. Exp.*, 2017, **25**, 17639.

33 S. Yu, X. Wu, K. Chen, B. Chen, X. Guo, D. Dai, L. Tong, W. T. Liu, and Y. Shen, *Optica*, 2016, **3**, 541.

34 R. Yan, D. Gargas and P. Yang, *Nat. Photonics,* 2009, **3**, 569.

35 K. L. Kompa and R. D. Levine, *Proc. Natl. Acad. Sci.*, 2001, **98**, 410.

36 B. Kolodziejczyk, Ch. H. Ng, X. Strakosas, G. G. Malliaras and B. Winther-Jensen, ***Mater. Horiz.***, 2018, **5**, 93.

37 G. Pawlik, A. C. Mitus, J. Mysliwiec, A. Miniewicz and J. G. Grote, *Chem. Phys. Lett.,* 2010, **484**, 321.

38 A. C. Mitus, G. Pawlik, A. Kochalska, J. Mysliwiec, A. Miniewicz and F. Kajzar, *Proc. SPIE,* 2007, **6646**, 664601.

39 H. You, H. Spaeth, V. N. L. Linhard and A. J. Steckl, *Langmuir*, 2009, **25**, 11698.

Published in Journal of Materials Chemistry C, doi: 10.1039/c8tc04677h (2018).40 Y. Kawabe, L. Wang, S. Horinouchi and N. Ogata, *Adv. Mater.*, 2000, **12**, 1281.

41 M. Dumont and A. E. Osman, *Chem. Phys.*, 1999, **245**, 437.

42 J. Mysliwiec, A. Szukalski, L. Sznitko, A. Miniewicz, K. Haupa, K. Zygadlo, K. Matczyszyn, J. Olesiak-Banska and M. Samoc, *Dyes Pigm.,* 2014, **102**, 63.

43 A. Szukalski, L. Sznitko, K. Cyprych, A. Miniewicz and J. Mysliwiec, *J. Phys. Chem. C,* 2014, **118**, 8102.

44 J. Mysliwiec, L. Sznitko, A. Szukalski, K. Parafiniuk, S. Bartkiewicz, A. Miniewicz, B. Sahraoui, I. Rau and F. Kajzar, *Opt. Mater.,* 2012, **34**, 1725.

45 L. Sznitko, J. Mysliwiec, K. Parafiniuk, A. Szukalski, K. Palewska, S. Bartkiewicz and A. Miniewicz, *Chem. Phys. Lett.,* 2011, **512**, 247.

46 A. Miniewicz, K. Palewska, J. Lipinski, R. Kowal and B. Swedek, *Mol. Cryst. Liq. Cryst.,* 1994, **253**, 41.

47 I. Papagiannouli, A. Szukalski, K. Iliopoulos, J. Mysliwiec, S. Couris and B. Sahraoui, *RSC Adv.,* 2015, **5**, 48363.

48 A. Szukalski, B. Sahraoui, B. Kulyk, C. A. Lazar, A.-M. Manea and J. Mysliwiec, *RSC Adv.*, 2017, **7**, 9941.

49 A. Szukalski, K. Haupa, A. Miniewicz and J. Mysliwiec, *J. Phys. Chem. C,* 2015, **119**, 10007.

50 A. Szukalski, A. Miniewicz, K. Haupa, B. Przybyl, J. Janczak, A. L. Sobolewski and J. Mysliwiec, *J. Phys. Chem. C,* 2016, **120**, 14813.

51 M. Poprawa-Smoluch, J. Baggerman, H. Zhang, H. P. A. Maas, L. De Cola and A. M. Brouwer, *J. Phys. Chem. A,* 2006, **110**, 11926.

52 H. Tian and J. Zhang, *Photochromic Materials: Preparation, Properties and Applications*, Wiley-VCH, John Wiley & Sons, Weinheim, Germany 2016.

53 X. Yao, T. Li, J. Wang, X. Ma and H. Tian, *Adv. Opt. Mater.*, 2016, **4**, 1322.

54 V. Cantatore, G. Granucci and M. Persico, *Phys. Chem. Chem. Phys.*, 2014, **16**, 25081.

55 L. Brzozowski and E. H. Sargent, *J. Mater. Sci.: Mater. Electron.*, 2001, **12**, 483.

56 R. L. Sutherland, *Handbook of Nonlinear Optics*, 2nd Ed., Marcel Dekker Inc. New York, USA, 2003.

57 R. W. Boyd, *Nonlinear Optics*, Academic Press, 3rd Ed., New York, 2008.

58 E. Hebda, M. Jancia, F. Kajzar, J. Niziol, J. Pielichowski, I. Rau and A. Tane, *Mol. Cryst. Liq. Cryst.*, 2012, **556**, 309.

59 F. Di Benedetto, E. Mele, A. Camposeo, A. Athanassiou, R. Cingolani and D. Pisignano, *Adv. Mater.*, 2008, **20**, 314.

60 D. Bleger and S. Hecht, *Angew. Chem. Int. Ed.,* 2015, **54**, 11338.

61 J. M. Hales, J. Matichak, S. Barlow, S. Ohira, K. Yesudas, J.-L. Brédas, J. W. Perry and S. R. Marder, *Science*, 2010, **327**, 1485.
12

62  Z. Li, Y. Liu, H. Kim, J. M. Hales, S.-H. Jang, J. Luo, T. Baehr-Jones, M. Hochberg, S. R. Marder, J. W. Perry and A. K.-Y. Jen, *Adv. Mater.*, 2012, **24**, OP326.

63  Z. Li, T. R. Ensley, H. Hu, Y. Zhang, S.-H. Jang, S. R. Marder, D. J. Hagan, E. W. Van Stryland and A. K.-Y. Jen, *Adv. Optical Mater.*, 2015, **3**, 900.

64  M. Hoffmann, I. N. Papadopoulos and B. Judkewitz, *Opt. Lett.*, 2018, **43**, 22.

65  D. Li and Y. Xia, *Adv. Mater.,* 2004, **16**, 1151.

66  M. V. Kakade, S. Givens, K. Gardner, K. H. Lee, D. B. Chase and J. F. Rabolt, *J. Am. Chem. Soc.,* 2007, **129**, 2777.

67  S. Pagliara, M. S. Vitiello, A. Camposeo, A. Polini, R. Cingolani, G. Scamarcio and D. Pisignano, *J. Phys. Chem. C,* 2011, **115**, 20399.

68  M. Richard-Lacroix and C. Pellerin, *Macromolecules,* 2013, **46**, 9473.

# ELECTRONIC SUPPLEMENTARY INFORMATION

## All-optical switching in dye-doped DNA nanofibers

Adam Szukalski,[a,b] Maria Moffa,[b] Andrea Camposeo,[b] Dario Pisignano,[b,c] and Jaroslaw Mysliwiec [a]

[d.] *Faculty of Chemistry, Wroclaw University of Science and Technology, Wybrzeze Wyspianskiego 27, 50-370 Wroclaw, Poland*
[e.] *NEST, Istituto Nanoscienze-CNR, Piazza S. Silvestro 12, I-56127 Pisa, Italy*
[f.] *Dipartimento di Fisica, Università di Pisa, Largo B. Pontecorvo 3, I-56127 Pisa, Italy. E-mail: dario.pisignano@unipi.it*

## Section S1. Synthetic route and technical details of PY-*p*CN compound synthesis, DNA-CTMA functionalization and biofibers preparation

**PY-*p*CN.** The synthesis route of (*E*)-4-(2-(1-phenyl-4,5-dihydro-1*H*-pyrazol-3-yl)vinyl)benzonitrile (PY-*p*CN) was performed by the following steps, presented also in Fig. 1c. At first 4-(bromomethyl)benzonitrile was synthesized. 4-methylbenzonitrile (25 g, 0.21 mol) was dissolved in 150 cm$^3$ of CCl$_4$. Equimolar amount of N-Bromosuccinimide (NBS) and 0.3 g of azobisisobutyronitrile (AIBN) were added. The reaction mixture was refluxed for five hours and after cooling to 40°C it was filtered and the used solvent left to evaporate. The residue was chromatographed on silica gel with dichloromethane as eluent. Afterwards, only the first fraction was collected. The yield of this step of synthesis was 25 g (57 %). According to the following procedure, the final compound PY-*p*CN was obtained. The 4-(bromomethyl)benzonitrile from the abovementioned first step of synthesis (in quantity 0.4 g, 2 mmol) and triphenylphosphine (0.524 g, 2 mmol) were dissolved and boiled in dry benzene overnight. The resulting salt was filtered, washed with hot benzene and used without further purification. To the suspension of phosphonium salt in dry tetrahydrofuran (THF, 25 cm$^3$), under inert atmosphere at room temperature, sodium ethanolate (0.108 g, 2 mmol) was added. The mixture color became deep red and afterwards the solution was stirred for further 30 minutes. Subsequently the solution of 1-phenyl-4,5-dihydro-1*H*-pyrazole-3-carbaldehyde (0.348 g, 2 mmol) in dry THF (10 cm$^3$) was added drop wise and it was stirred overnight at temperature equal to 50°C. Then the solvent was evaporated and dichloromethane was added to the orange residue until it became homogenous. The final product was purified on silica gel with dichloromethane as eluent. Finally, it was crystallized from heptane. The reaction yield was 0.325 g (59.5 %).

The PY-*p*CN structure was confirmed by $^1$H NMR and infrared spectroscopy. $^1$H NMR analysis of PY-*p*CN compound was performed by using a Bruker Avance III (300 MHz) apparatus. $^1$H-NMR (CDCl$_3$, 300 MHz) δ (ppm): 7.61 (m, 2H),

7.47 (m, 2H), 7.29 (m, 2H), 7.05 (m, 2H), 6.87 (m, 1H), 6.66 (m, 1H), 3.90 (t, 1H), 3.69 (t, 1H), 3.12 (t, 1H), 2.91 (s, 1H), 2.56 (t, 1H).

Fourier-transform infrared (FTIR) spectra were recorded in range 4000-400cm$^{-1}$ using KBr discs on a Bruker Vertex 70 spectrometer. A complete list of the characteristic vibrations coming from the functional groups for this compound is presented below:

IR (KBr) ν(cm$^{-1}$): 3065, 2969, 2926, 2866, 2225, 1646, 1598, 1533, 1503, 1475, 1455, 1417, 1400, 1366, 1349, 1311, 1297, 1283, 1242, 1223, 1208, 1172, 1146, 1125, 1067, 1033, 1017, 999, 974, 964, 956, 948, 882, 865, 816, 775, 753, 697, 692, 664, 586, 554, 531, 506, 468, 431.

In FTIR spectra, all of the vibrations characteristic for the investigated pyrazoline derivative (such as: C-H aromatic, C-H aliphatic, C-N, C=N) were observed. Few characteristic vibrations should be distinguished, i.e. for the -C=C- group vibration localized close to 1646 cm$^{-1}$ or for the nitrile one (-C≡N) around 2225 cm$^{-1}$.

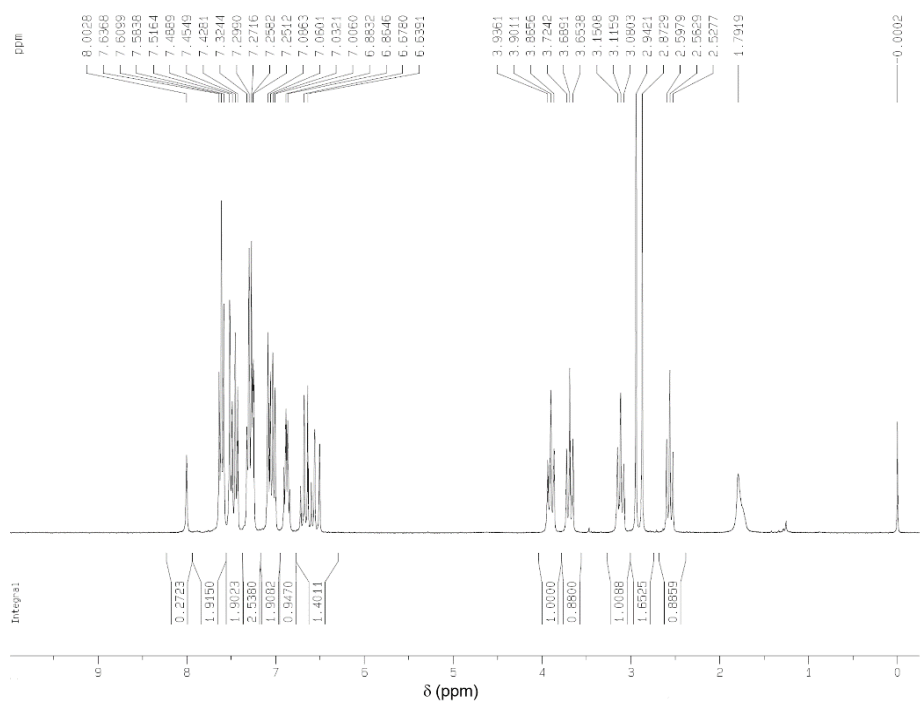

**Figure S1**. $^1$H NMR spectrum of PY-$p$CN (CDCl$_3$, 300 MHz).

The photoluminescence (PL) spectrum of PY-*p*CN is shown in Figure S2. The spectrum was measured by using Hitachi F-4500 FL spectrophotometer. Applied scan speed was set at: 240 nm/min.

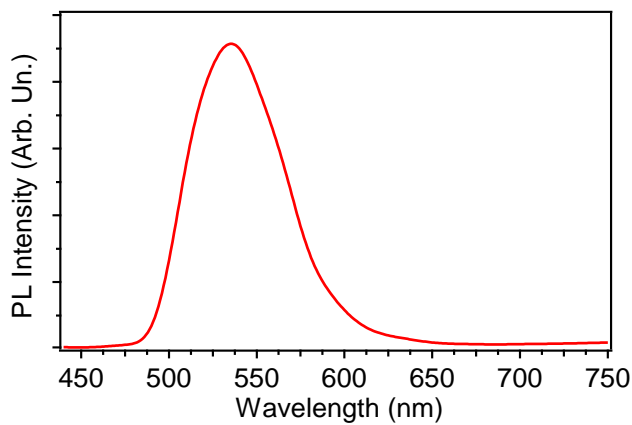

**Figure S2**. Photoluminescensce spectrum of PY-*p*CN.

**DNA-CTMA.** DNA functionalization by cetyltrimetylammonium chloride (CTMA) surfactant was carried out according to the methodology presented by Heckman and described in the literature.[S1] At the beginning the proper mass of DNA was weighted $m_{DNA}$ = 0.67 g and then dissolved in 200 mL of deionized water. The solution was warmed up ($T$ = 40 ºC) and stirred till the moment when no parts of biopolymeric matrix were visible. The same portion of CTMA surfactant was taken and dissolved in deionized water with final concentration equal to $c_{CTMA}$ = 4 g/dm$^3$. The latter solution was added drop wise to water mixture with DNA, forming in this way DNA-CTMA precipitate. The reaction mixture was stirred for two hours at room temperature and the final heterogeneous mixture was filtered through cellulose filter. The used filter was rinsed with deionized water to ensure that there is no undissolved or not reacted free CTMA left in pellet. Precipitate collected from the filter was dried in oven for 24 h at higher temperature ($T$ = 40 ºC) to evaporate the solvent.

## Section S2. Electrospun nanofiber size distribution

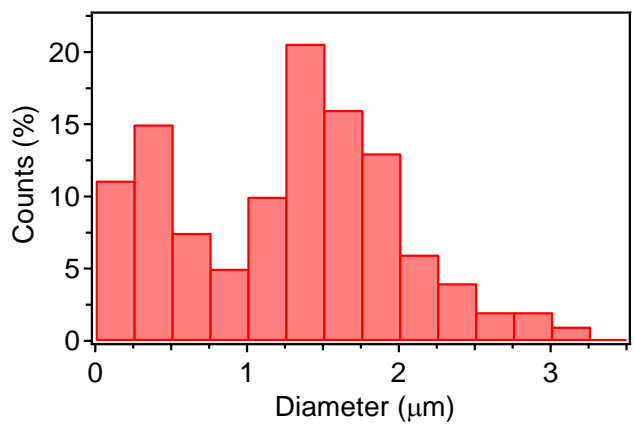

**Figure S3**. Distribution of the diameters of the DNA-CTMA electrospun nanofibers doped with PY-*p*CN.

## Section S3. All-optical switching measurements

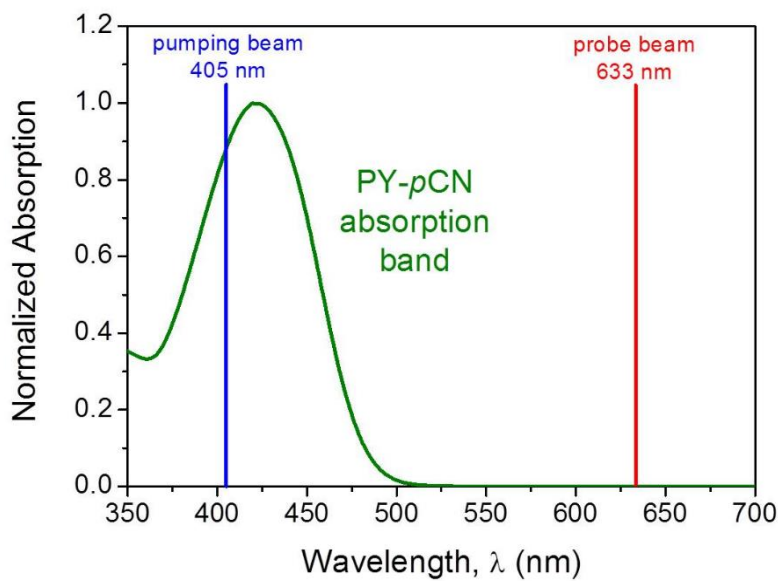

**Figure S4.** PY-*p*CN absorption in solid state and used pump/probe laser lines.

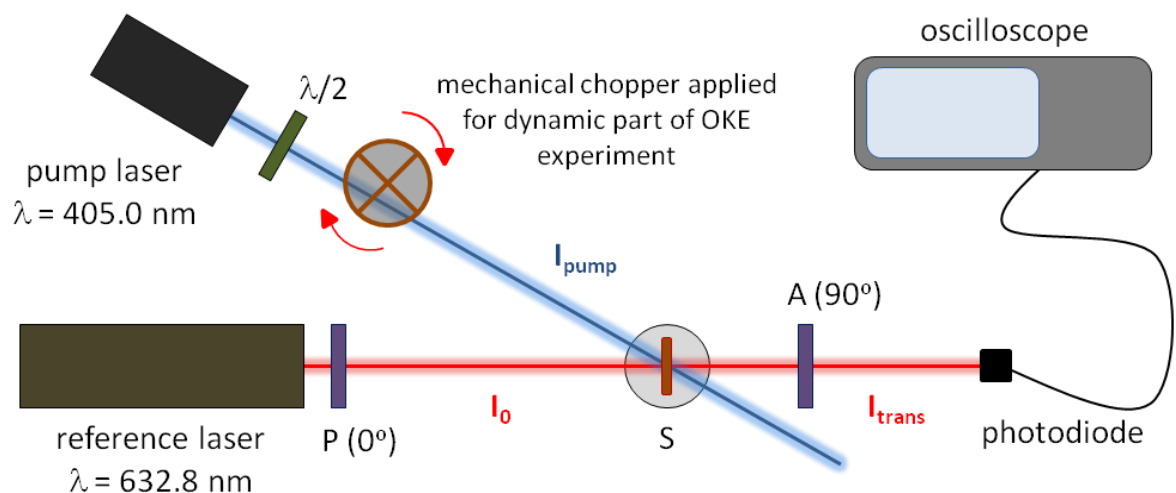

**Figure S5.** Scheme of the used experimental set-up.

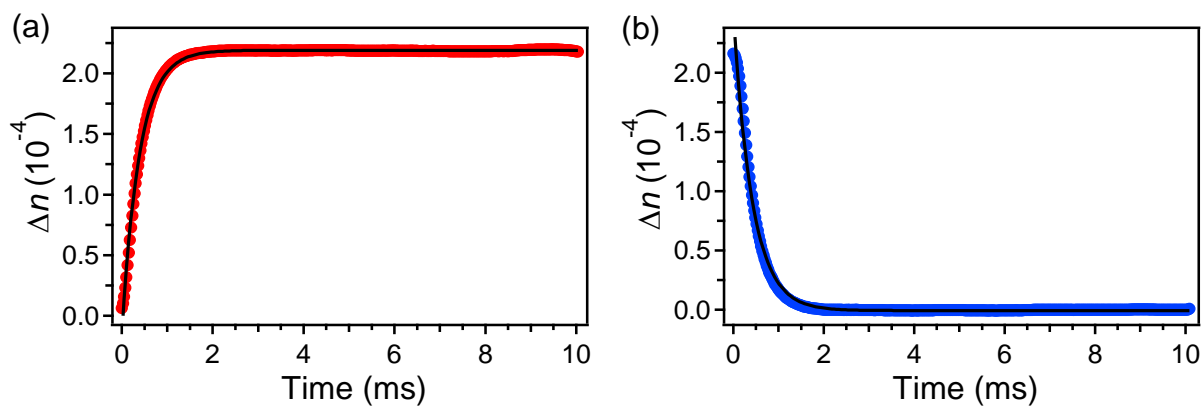

**Figure S6.** Exemplary behaviour of the increase (a) and decay (b) of signals in long-term photo-ordering for PY-*p*CN/DNA-CTMA fibers. $\tau_{inc}$ and $\tau_{dec}$, characteristic times for signal increase and decay, are 0.39 ms and 0.42 ms, respectively. The continuous lines are fit to the data by single exponential function. Excitation intensity from the pump beam: 750 mW/cm$^2$.

**References**

S1    E. M. Heckman, J. A. Hagen, P. P. Yaney, J. G. Grote and F. K. Hopkins, *Appl. Phys. Lett.,* 2005, **87**, 211115.